\documentclass[conference]{IEEEtran}
\usepackage{amsmath,amssymb}
\usepackage{algorithm}
\usepackage{algorithmic}
\usepackage{graphicx}
\usepackage{xcolor}
\usepackage{fancyhdr}
\IEEEoverridecommandlockouts

\begin{document}

\title
{Service Function Chain Routing in LEO Networks Using Shortest-Path Delay Statistical Stability}

\author{
    Li Zeng$^{*}$, Zixin Wang$^{*}$, Yuanming Shi$^{\dag}$, Khaled B. Letaief$^{*}$
    \thanks{The work of Yuanming Shi was supported in part by the National Natural Science Foundation of China under Grant 62522117. 
    This work was supported in part by the Hong Kong Research Grants Council under the Area of Excellence (AoE) Scheme Grant No. AoE/E-601/22-R.
    }
    \\[0.5ex]
    \normalsize{$^{*}$Dept. of Electronic and Computer Engineering, The Hong Kong University of Science and Technology, Hong Kong}\\
    \normalsize{$^{\dag}$School of Information Science and Technology, ShanghaiTech University, Shanghai, China}\\
    \normalsize{Emails: \{lzengan@connect.ust.hk; eewangzx@ust.hk; shiym@shanghaitech.edu.cn; eekhaled@ust.hk\}}
}

\maketitle

{\color{black}
\begin{abstract}
Low Earth orbit (LEO) satellite constellations have become a critical enabler for global coverage, utilizing numerous satellites orbiting Earth at high speeds. By decomposing complex network services into lightweight service functions, network function virtualization (NFV) transforms global network services into diverse service function chains (SFCs), coordinated by resource-constrained LEOs.
However, the dynamic topology of satellite networks, marked by highly variable inter-satellite link delays, poses significant challenges for designing efficient routing strategies that ensure reliable and low-latency communication. Many existing routing methods suffer from poor scalability and degraded performance, limiting their practical implementation.
To address these challenges, this paper proposes a novel SFC routing approach that leverages the statistical properties of network link states to mitigate instability caused by instantaneous modeling in dynamic satellite networks. Through comprehensive simulations on end-to-end shortest-path propagation delays in LEO networks, we identify and validate the statistical stability of multi-hop routes.
Building on this insight, we introduce the Stability-Aware Multi-Stage Graph Routing (SA-MSGR) algorithm, which incorporates pre-computed average delays into a multi-stage graph optimization framework. Extensive simulations demonstrate the superior performance of SA-MSGR, achieving significantly lower and more predictable end-to-end SFC delays compared to representative baseline strategies.
\end{abstract}
}

\section{Introduction}
\label{sec:intro}
{\color{black}The rapid growth of satellite constellations has become a vital enabler for achieving ubiquitous connectivity, providing seamless global coverage through thousands of low Earth orbit (LEO) satellites orbiting at high speeds \cite{10605604}.
For instance, Starlink’s constellation of over 7,000 LEO satellites delivers global internet services with download speeds ranging from 25 to 220 Mbps \cite{Starlink2025}.
However, unlike terrestrial and near-ground networks, which possess substantial computational capabilities, individual satellites in LEO constellations are constrained by limited computational resources. This limitation poses significant challenges for the implementation of advanced network services, such as military remote sensing and long-distance maritime video communication.
Network function virtualization (NFV) can address these challenges by decomposing advanced network services into a series of lightweight service functions, which can then be flexibly mapped onto distributed LEOs within satellite networks, and each network service can thus be represented as a service function chain (SFC) \cite{wang2020sfc}.
Nevertheless, the dynamic topology of satellite constellations, driven by the high mobility of LEOs, results in frequent changes in inter-satellite link (ISL) states, acting as a performance bottleneck for delivering reliable and low-latency network services. Consequently, developing efficient function-routing algorithms among LEOs becomes imperative.}

{\color{black} Many of the existing research has proposed modeling the dynamic LEO network using a time-expanded graph (TEG) \cite{yang2021maximum, zhang2024quantum}. The core concept of TEG involves leveraging the temporal correlations in the satellite network topology and representing the network's evolution over a specified time window as a static graph, wherein nodes and edges are replicated across discrete time slots.
This approach has been utilized, for instance, to find maximum flow paths to minimize the delay considering service constraints \cite{yang2021maximum} and further accelerated using quantum-assisted methods \cite{zhang2024quantum}.
However, the size of the TEG scales as the product of the network size and the time window length, leading to low scalability as the network scale increases. Furthermore, the method relies on multiple hyperparameters, such as the time window length, which necessitate careful tuning \cite{guo2024enhanced} and reduce reliability thereafter.}

To achieve scalable routing decisions, snapshot-based methods have emerged as a promising approach. Such methods generate routing decisions based on the quasi-static network topology to schedule SFC tasks \cite{wang2020sfc, zhang2022space}, and optimize delay-sensitive VNF deployment \cite{xu2024delay}, capitalizing on their lower complexity compared to time-aware methods.
{\color{black}However, under the rapidly evolving satellite topology, routing decisions derived from instantaneous network states may quickly become suboptimal or infeasible due to topology changes, thereby compromising reliability and latency requirements.
Moreover, ensuring route validity in a dynamic environment requires frequent recomputation based on updated network states, imposing a significant computational burden, especially on resource-constrained LEOs.}

{\color{black}
In this paper, we propose a novel SFC routing approach that harnesses the statistical properties of network link states to alleviate instability arising from instantaneous modeling in dynamic satellite networks.
Specifically, we first demonstrate the statistical stability of end-to-end shortest-path transmission delays in LEO networks through systematic simulations. Our findings reveal that the majority of satellite pairs exhibit minimal relative fluctuations, with stability further enhanced in multi-hop SFC paths. This improvement is attributable to the reduced coefficient of variation in longer chain lengths. 
Building on this observation, we propose the Stability-Aware Multi-Stage Graph Routing (SA-MSGR) algorithm. This approach employs pre-computed average shortest-path transmission delays as static weights within a multi-stage graph framework designed for SFC requests. By decoupling the complexity of dynamic optimization from online path computation, the SA-MSGR algorithm enables efficient and scalable routing decisions. 
The key contributions are summarized as follows:}
\begin{enumerate}
\item Systematic experimental validation of the statistical stability of pairwise and multi-hop shortest-path transmission delays in representative LEO constellations.
\item Development of the SA-MSGR algorithm, which leverages delay stability within a multi-stage graph framework to achieve efficient SFC routing with low online complexity.
\item Comprehensive performance evaluation demonstrating that SA-MSGR can reduce end-to-end SFC delays compared to existing baseline heuristics.
\end{enumerate}

The remainder of this paper is organized as follows. Section~\ref{sec:system_model} details the system model, including the LEO network architecture and SFC request definitions. Section~\ref{sec:stability} presents the experimental investigation into the statistical stability of shortest-path delays. Section~\ref{sec:prob_form} formulates the delay minimization problem and introduces the proposed SA-MSGR algorithm. Section~\ref{sec:sim} evaluates the performance of SA-MSGR through simulations. Finally, Section~\ref{sec:conclusion} concludes the paper.

\section{System Model}
\label{sec:system_model}

We use the most widely adopted Walker Delta configuration \cite{wang2022capacity} as the underlying LEO network architecture. This configuration is typically denoted by $T/P/F$, where $T$ is the total number of satellites in the constellation, $P$ is the number of equally spaced orbital planes, $N = T/P$ is the number of satellites per orbital plane, and $F$ is the relative phasing parameter for satellites between adjacent planes.
Let the set of all satellites be denoted as
\vspace{-1mm}
\begin{equation}
\vspace{-1.5mm}
    \mathcal{S} = \{ s_{i,j} \mid i = 1, 2, \ldots, P; \; j = 1, 2, \ldots, N \},
\end{equation}
where $s_{i,j}$ represents the $j$-th satellite in the $i$-th orbital plane. The total number of satellites is $|\mathcal{S}| = T = P \times N$.

We assume that all $P$ orbital planes share the same altitude and inclination, which are distributed uniformly by their right ascension of the ascending node, with an angular separation of $2\pi/P$ between adjacent planes. In each plane $i$, the $N$ satellites are evenly spaced along the orbit, meaning the difference in mean anomaly (an angle representing position along the orbit) between consecutive satellites $s_{i,j}$ and $s_{i,j+1}$ is $2\pi/N$\footnote{Index arithmetic for planes and satellites (e.g., $i\pm 1$, $j\pm 1$) is implicitly modulo $P$ and $N$ respectively, where applicable.}. The Walker Delta phasing parameter $F$ introduces a specific offset $2\pi F / T$ in mean anomaly between corresponding satellites in adjacent planes ($i$ and $i+1$), ensuring a structured relative positioning across planes. Connectivity between satellites is established via bidirectional laser ISLs. We use $\mathcal{E}(t)$ to denote the set of all ISLs at time $t$. The connectivity pattern typically follows these rules:
\begin{itemize}
    \item \textbf{Intra-plane connectivity:} Satellites within the same orbital plane maintain stable connections with their immediate neighbors, i.e., $s_{i,j}$ is typically connected to $s_{i,j\pm1}$.
    \item \textbf{Inter-plane connectivity:} Satellites in adjacent orbital planes (e.g., plane $i$ and plane $i \pm 1$) establish connections if they are sufficiently close and operational constraints are met. More specifically, each satellite $s_{i,j}$ typically attempts to connect with its geometrically nearest neighbor in the preceding and succeeding plane $i\pm1$. Besides, such a potential inter-plane ISL is only activated if it satisfies the requirements: 1) the line-of-sight between the two satellites must not be obstructed by the Earth; 2) the relative velocity between the pair must not exceed a threshold $V_{\text{th}}$, as excessively high relative speeds can compromise link stability and tracking accuracy. These conditions ensure that active inter-plane links are physically viable and maintainable.
\end{itemize}
Typically, for most satellites, four stable ISL connections exist: two intra-plane links connecting neighboring satellites and two inter-plane links to adjacent orbital planes. 

\subsection{Laser Inter-Satellite Link Model}
As in \cite{shang2025channel}, we characterize each active laser ISL $e = (u, v) \in \mathcal{E}(t)$ at time $t$ with its propagation delay, which is dominant in the transmission latency. Let $L_e(t)$ denote the Euclidean distance between the connected satellites $u$ and $v$ at time $t$. The propagation delay on link $e$ at time $t$ is given by:
$
d_e(t) = \frac{L_e(t)}{c},
$
where $c$ is the speed of light in a vacuum. Since satellites are constantly moving, $L_e(t)$ and consequently $d_e(t)$ are time-varying, especially for inter-plane links. Note that the Laser ISLs are typically full-duplex and offer very high channel capacity $C_e$, potentially reaching 100 Gbps or more \cite{brashears2024achieving}. Therefore, for typical data packet sizes $S$ envisioned in SFC scenarios, the transmission delay component $t_e = S / C_e$ {\color{black}can be omitted} compared to the propagation delay $d_e(t)$ \cite{ekici2001distributed}. 

While establishing an ISL requires precise acquisition, tracking, and pointing (ATP) \cite{bhattacharjee2023laser}, the intra-plane links and inter-plane links outside polar regions generally exhibit sufficient stability for long-term connection \cite{shang2025channel}. However, it is crucial to note that the stability of link existence does not imply constant link delay. The distance $L_e(t)$ still varies. Although individual link delays $d_e(t)$ fluctuate, our work will later investigate the statistical stability of the end-to-end shortest path delays composed of multiple such links.

\subsection{Service Function Chain Task Request}
\begin{figure}[t]
    \centering
    \includegraphics[width=1.0\linewidth]{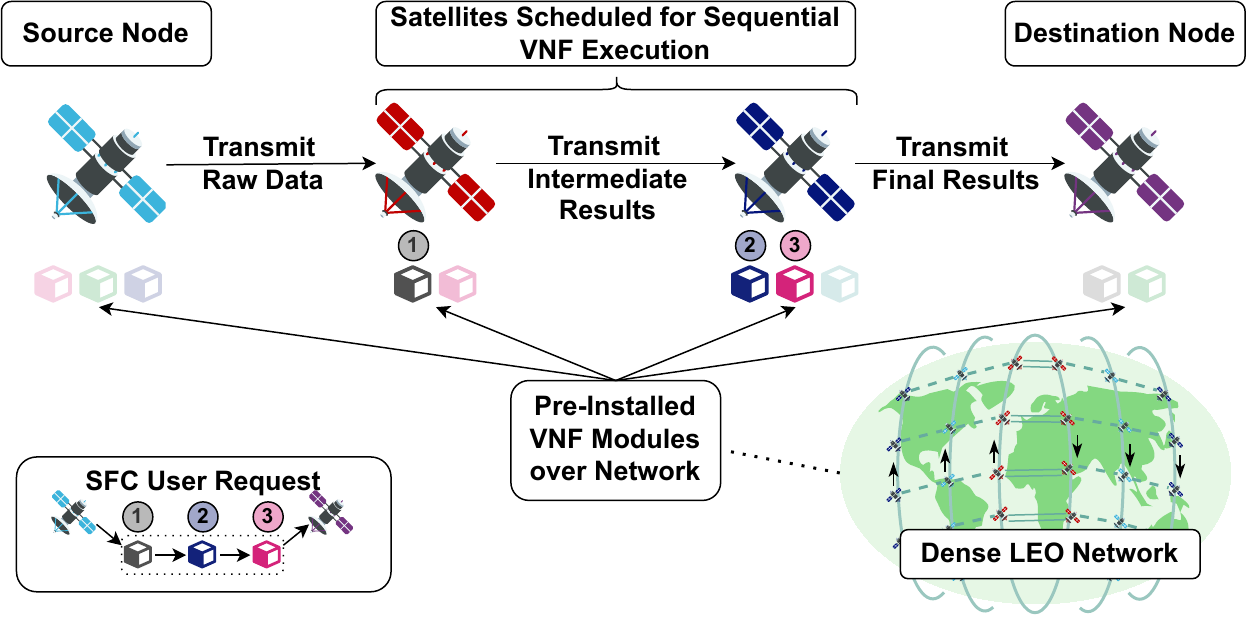}
    \caption{Illustration of the system model. An SFC user request specifies an ordered sequence of VNFs (e.g., 1, 2, 3) between a source and destination. This logical chain is mapped onto the physical LEO network, where satellites host corresponding pre-installed VNF modules.}
    \label{fig:sysmdl}
    \vspace{-5mm}
\end{figure}

Within the LEO network, we assume each satellite $s \in \mathcal{S}$ may be pre-deployed with a set of VNF modules. Let $\mathcal{V}(s) \subseteq \mathcal{V}$ denote the set of VNFs available on satellite $s$, where $\mathcal{V}$ is the universal set of all possible VNFs.
A SFC request specifies an ordered sequence of VNFs that must be executed, processing data originating from a source satellite and delivering the final result to a destination satellite. Formally, an SFC request $r$ is defined as a tuple:
$
r = \Big( s_{\text{src}},\, s_{\text{dst}},\, \mathbf{f} \Big),
$
where $s_{\text{src}} \in \mathcal{S}$ is the source satellite holding the initial input data, $s_{\text{dst}} \in \mathcal{S}$ is the destination satellite to receive the final result and $\mathbf{f} = [ f_1, f_2, \ldots, f_M ]$ is the ordered list of $M$ VNFs ($f_k \in \mathcal{V}$) to be executed sequentially.
Each VNF $f_k$ in the sequence has specific characteristics: \( D_{\text{in}, f_k} \) be the input data size required by VNF \( f_k \), \( D_{\text{out}, f_k} \) be the output data size produced by VNF \( f_k \), and $C_{f_k}$ is the computational complexity of VNF $f_k$, measured in floating-point operations (FLOPs).
Each satellite $s \in \mathcal{S}$ possesses a processing unit with a computational capacity $F_s$ (FLOPs per second). The processing delay incurred when executing VNF $f_k$ on satellite $s$ (assuming $f_k \in \mathcal{V}(s)$) is:
\vspace{-1mm}
\begin{equation}\label{eq:comp_delay}
    \vspace{-3mm}
    D^{\text{cp}}_{s,f_k} = \frac{C_{f_k}}{F_s}.
    \vspace{0mm}
\end{equation}
We note that a VNF $f_k$ can only be processed if it is available on the chosen satellite, i.e., $f_k \in \mathcal{V}(s_k)$. The system model is illustrated in Fig. \ref{fig:sysmdl}.

\subsection{SFC Routing Path and End-to-End Delay}
To fulfill an SFC request $r = (s_{\text{src}}, s_{\text{dst}}, \mathbf{f})$, a sequence of satellites must be chosen to execute the VNFs in the specified order. A feasible execution path for request $r$ is represented as:
$
\pi(r) = \{ s_0, s_1, \ldots, s_{M+1} \},
$
where $s_0 = s_{\text{src}}$ is the source satellite, $s_{M+1} = s_{\text{dst}}$ is the destination satellite. Additionally, for each $k = 1, \ldots, M$, satellite $s_k \in \mathcal{S}$ is chosen to execute VNF $f_k$, and must satisfy the deployment constraint $f_k \in \mathcal{V}(s_k)$.
The total end-to-end delay for executing the SFC along path $\pi(r)$ depends on both the computational delays at the chosen satellites and the transmission delays between consecutive satellites in the path. Since the network topology $\mathcal{G}(t) = (\mathcal{S}, \mathcal{E}(t))$ and link propagation delays $d_e(t)$ are time-varying, the transmission delay between any two satellites $u, v \in \mathcal{S}$ also varies with time. 

To minimize the total delay, we assume that transmissions between satellite pairs adhere to the shortest path, which is also the fundamental strategy adopted by modern data routing protocols. Let $D^{\text{tx}}(u, v, t)$ denote the minimum total propagation delay along a path connecting satellite $u$ to satellite $v$ at time $t$. This is the shortest path transmission delay found by summing the instantaneous link delays $d_e(t)$ for $e \in \mathcal{E}(t)$ over the path in $\mathcal{G}(t)$ with the minimum total sum (e.g., using Dijkstra's algorithm on $\mathcal{G}(t)$ with weights $d_e(t)$).
\begin{equation}
    D^{\text{tx}}(u, v, t) = \min_{p \in \mathcal{P}_{u,v}(t)} \sum_{e \in p} d_e(t),
    \vspace{-1.5mm}
\end{equation}
where $\mathcal{P}_{u,v}(t)$ is the set of all paths from $u$ to $v$ in $\mathcal{G}(t)$. 

Consequently, all link delays to be summed are calculated based on the network state at time $ t $, denoted as $ d_e(t) $.
The total instantaneous end-to-end delay for executing SFC request $r$ along path $\pi(r)$ at time $t$ is the sum of the computation delays for all VNFs and the shortest path transmission delays between consecutive satellites in the execution sequence:
\vspace{-1mm}
\begin{equation}\label{eq:total_inst_delay}
\vspace{-1.5mm}
    D(\pi(r), t) = \sum_{k=1}^{M} D^{\text{cp}}_{s_k, f_k} + \sum_{k=0}^{M} D^{\text{tx}}(s_k, s_{k+1}, t_k),
\end{equation}
where $ t_k $ is the starting time of transmission from $s_k$ to $s_{k+1}$, depending on the previous transmission and computation process. This instantaneous delay $D(\pi(r), t)$ fluctuates over time due to the $D^{\text{tx}}(s_k, s_{k+1}, t_k)$ term, which depends on the dynamic network state $\mathcal{G}(t)$. 

To develop robust routing strategies, it is essential to understand the statistical nature of the fluctuating component. Specifically, 
{\color{black}We hypothesize that:
1) A significant portion of satellite pairs might exhibit statistically stable shortest path delays (i.e., low variance relative to the mean).
2) The stability of the total transmission delay along an SFC path might increase as the number of hops ($M+1$) increases, potentially due to averaging effects. 
These hypotheses will be tested through a detailed experimental investigation in the Section \ref{sec:stability}, with the findings serving as the empirical basis for the proposed routing algorithm presented in Section \ref{sec:prob_form}.}

\section{The Shortest-Path Delay Stability Property}
\label{sec:stability}

In this section, we demonstrate through simulations the statistical stability of $D^{\text{tx}}(u, v, t)$ and $\sum_{k=0}^{M} D^{\text{tx}}(s_k, s_{k+1}, t_k)$,  a key property leveraged to tackle this routing challenge. In the simulation, we use Walker Delta configurations with 12 orbital planes, 30 satellites per plane, 550 km altitude, and 53 degrees inclination.
Network topology is updated every $\Delta t = 1$ seconds over one orbital period.

\subsection{Pairwise Shortest-Path Delay Stability}
\label{subsec:pairwise_stability}
We first analyze the stability over time of the shortest path propagation delay $D^{\text{tx}}(u, v, t)$ for individual satellite pairs. For each pair $(u, v)$ in the network, we use the coefficient of variation (CV) as the metric, defined as
\begin{equation}
    {CV}_{uv} = {\sigma_{uv}}/{\mu_{uv}},
\end{equation}
where $\sigma_{uv}$ is the standard deviation and $\mu_{uv}$ is the mean of the time series $\{D^{\text{tx}}(u, v, t) \mid t = 0, \Delta t, \ldots, T_{\text{orbit}}\}$. The CV provides a normalized measure of delay variability, where a lower CV indicates higher stability.

The stability distribution result is summarized in Table \ref{tab:cv_summary_transposed}. We observe that a large fraction of pairs exhibit low CV values. Specifically, $70\%$ of the satellite pairs have a CV less than 0.2, and $90\%$ have a CV less than 0.3, indicating a significant degree of shortest path delay stability despite the underlying network dynamics. This confirms our first hypothesis.

\begin{table}[t]
    \centering
    \caption{Stability of Pairwise Shortest Path Delays: CV Thresholds}
    \label{tab:cv_summary_transposed}
    \vspace{-2mm}
    \begin{tabular}{lccc}
    \hline
    \textbf{Pairs with CV $\le$ Threshold (\%)} & 50    & 70    & 90    \\
    \textbf{CV Threshold}                       & 0.125 & 0.198 & 0.295 \\
    \hline
    \end{tabular}
    \vspace{-3mm}
\end{table}
\begin{table}[t]
    \centering
    \caption{Average CV of Path Transmission Delay vs. SFC Length (M)}
    \label{tab:avg_cv_vs_m}
    \vspace{-2mm}
    \begin{tabular}{lccccc}
    \hline
    \textbf{SFC Length (M)} & 1     & 2     & 5     & 10    & 20    \\
    \textbf{Average CV}     & 0.099 & 0.079 & 0.053 & 0.041 & 0.028 \\
    \hline
    \end{tabular}
    \vspace{-3mm}
\end{table}

\subsection{SFC Path Total Propagation Delay Stability}
\label{subsec:path_stability}
Next, we investigate the stability of the \textit{total} transmission delay along a multi-hop SFC path. Our interest is whether the aggregation of delays along the path leads to increased stability compared to individual pairwise delays. We generate $N_{\text{paths}} = 500$ random paths $\pi = \{s_0, s_1, \ldots, s_{M+1}\}$ for different SFC lengths $M$ (e.g., $M = 1, 2, 5, 10, 20$). For each generated path $\pi$ and each time step $t$, we calculate the total instantaneous transmission delay:
$
D^{\text{tx}}_{\text{path}}(\pi, t) = \sum_{k=0}^{M} D^{\text{tx}}(s_k, s_{k+1}, t_k).
$
For each path $\pi$, we compute the coefficient of variation $CV_{\pi} = \sigma_{\pi} / \mu_{\pi}$ of its total transmission delay time series. 
The results are shown as follows:
\begin{itemize}
    \item \textbf{Stability Distribution vs. Path Length:} Fig. \ref{fig:cv_cdf_path} presents the CDF of the path CVs ($CV_{\pi}$), grouped by the SFC length $M$. The curves clearly shift towards the top-left as $M$ increases. This indicates that for longer SFC paths, a significantly larger fraction exhibits very low CV values (high stability).
    \begin{figure}[t]
        \centering
        \includegraphics[width=0.9\linewidth]{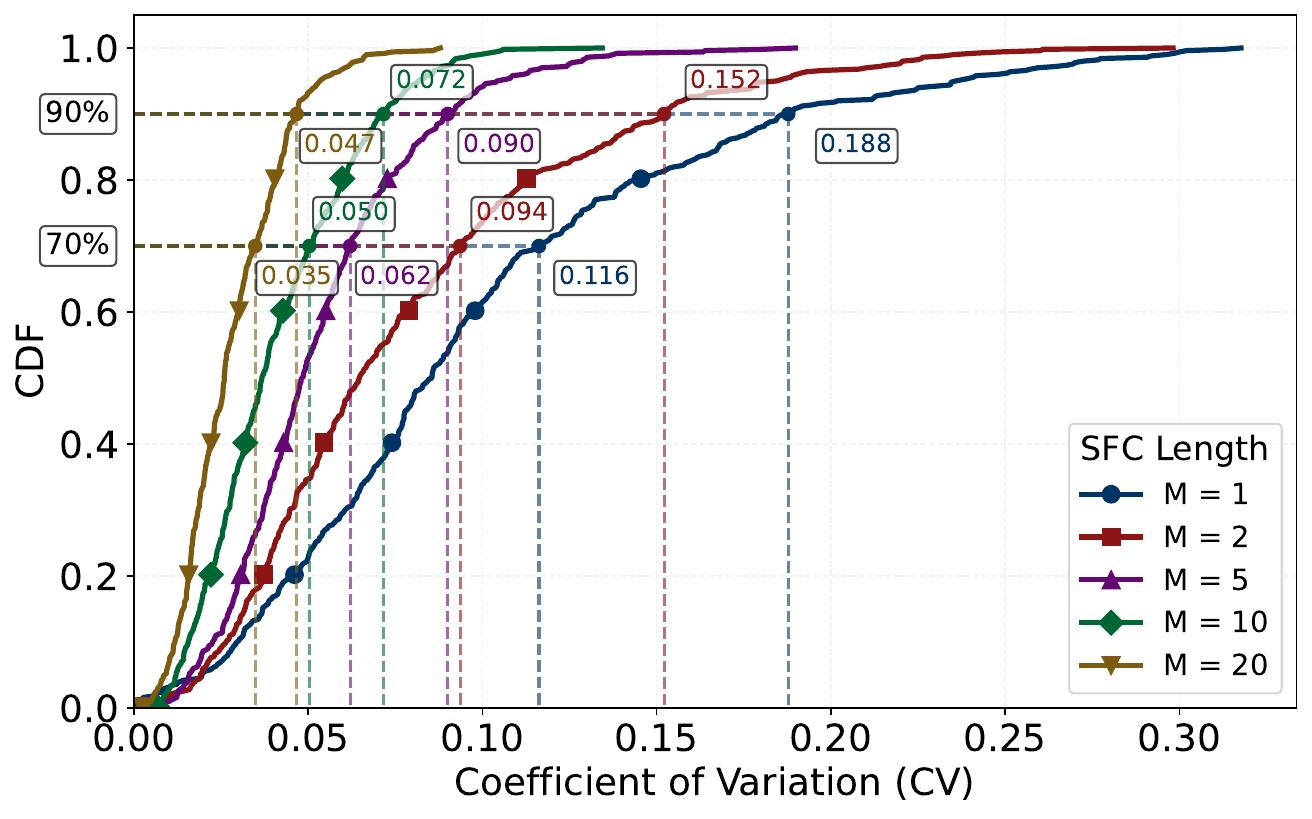}
        \vspace{-3mm}
        \caption{CDF of CV for path transmission delays with different lengths ($M$).}
        \label{fig:cv_cdf_path}
        \vspace{-3mm}
    \end{figure}
    \item \textbf{Average Stability Trend:} Table~\ref{tab:avg_cv_vs_m} lists the average CV (averaged over all $N_{\text{paths}}$ paths for a given $M$) as a function of the SFC length $M$. The average CV demonstrates a clear decreasing trend as $M$ increases, falling from 0.099 for $M=1$ to 0.028 for $M=20$, confirming that longer paths tend to be relatively more stable in terms of total transmission delay.
\end{itemize}
These results strongly support our second hypothesis. The aggregation of shortest-path delays along a multi-hop path tends to smooth out individual fluctuations, thereby enhancing the stability of the total transmission component of the SFC delay. This stabilizing effect becomes increasingly pronounced for longer chains, particularly in scenarios involving large-scale networks and highly complex service processes where the length of the service chain is substantial.

These findings have crucial implications for SFC routing. They provide strong empirical justification for approximating the complex, time-varying shortest path transmission delay $D^{\text{tx}}(u, v, t)$ with its pre-calculated average $\bar{D}^{\text{tx}}(u, v)$. In the next section, we will introduce the formulated SFC delay minimization problem and propose our method.

\section{Problem formulation and the Proposed SA-MSGR Algorithm}
\label{sec:prob_form}
{\color{black}To begin with}, we first establish the mathematical formulation of the delay minimization problem and then elaborate the proposed SA-MSGR method. Specifically, given an SFC request 
$
r = ( s_{\text{src}}, s_{\text{dst}}, \mathbf{f} = [f_1, \ldots, f_M] )
$ 
at time $t$, the primary goal is to determine the sequence of satellites 
$
\pi(r) = \{ s_0, s_1, \ldots, s_M, s_{M+1} \}
$ 
(where $s_0=s_{\text{src}}$, $s_{M+1}=s_{\text{dst}}$) that will host the VNFs $f_1, \ldots, f_M$ such that the overall end-to-end delay is minimized. 
As established before, the instantaneous end-to-end delay $D(\pi(r), t)$, i.e., Eq. (\ref{eq:total_inst_delay}) fluctuates over time due to the dynamic nature of the shortest path transmission delays $D^{\text{tx}}(s_k, s_{k+1}, t)$.
The optimization problem is then formulated as:
\begin{equation}\label{eq:avg_delay_obj}
    \begin{aligned}
        \operatorname*{minimize}_{\pi(r)} \quad &  D(\pi(r), t) \\
        \text{subject to} \quad & s_0 = s_{\text{src}},\quad s_{M+1} = s_{\text{dst}}, \\
        & f_k \in \mathcal{V}(s_k), \quad \forall\, k = 1,\ldots,M.
    \end{aligned}
\end{equation}
The critical challenge of this problem lies in the varying transmission delay brought by the great satellite mobility, i.e., a route computed based on a snapshot of the network topology may quickly become outdated; the optimal path under one quasi-static assumption may turn suboptimal, or even infeasible, as the network evolves.

However, building upon the delay stability properties validated in Section \ref{sec:stability}, we can propose an efficient SFC routing algorithm tailored for dynamic LEO networks. The core challenge, optimizing Eq. (\ref{eq:avg_delay_obj}), involves the time-varying transmission delays. Our approach leverages the observed stability (low CV) to justify approximating the expected delay with its pre-computed time average $\bar{D}^{\text{tx}}(u, v)$, obtained via offline simulation. This stability-aware multi-stage graph routing (SA-MSGR) method converts the dynamic problem into a static shortest path problem over a constructed multi-stage graph, sidestepping the complexities of TEG and the limitations of snapshot approaches.

\subsection{Multi-Stage Graph Construction} 
\begin{figure}[t]
    \centering
    \includegraphics[width=1.0\linewidth]{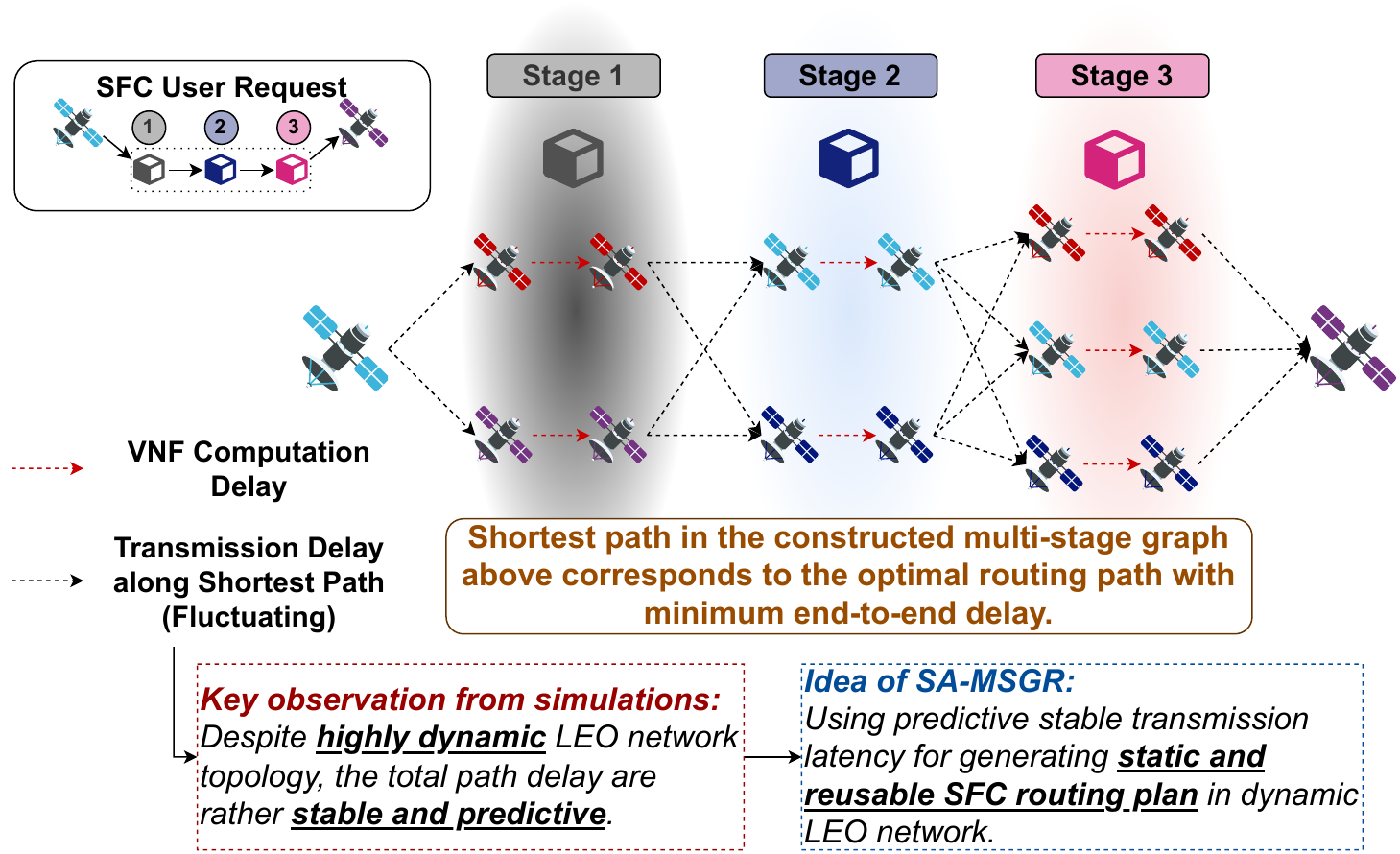}
    \caption{The proposed SA-MSGR method.}
    \label{fig:msg}
    \vspace{-5mm}
\end{figure}

We employ a Multi-Stage Graph (MSG) $\mathcal{G}_{\text{MSG}} = (\mathcal{V}_{\text{MSG}}, \mathcal{E}_{\text{MSG}})$ tailored to the specific SFC request $r = (s_{\text{src}}, s_{\text{dst}}, \mathbf{f}=[f_1, \ldots, f_M])$, as depicted in Fig. \ref{fig:msg}.

\begin{itemize}
    \item \textbf{Nodes $\mathcal{V}_{\text{MSG}}$:} The MSG comprises $M+2$ stages.
        \begin{itemize}
            \item Stage 0: Source node $N_{\text{src}}$ (representing $s_{\text{src}}$).
            \item Stage $k \in [1, M]$: Corresponds to VNF $f_k$. Contains nodes ($N_{s,k}^{\text{in}}, N_{s,k}^{\text{out}}$) for each satellite $s$ if $f_k \in \mathcal{V}(s)$.
            \item Stage $M+1$: Destination $N_{\text{dst}}$ (representing $s_{\text{dst}}$).
        \end{itemize}
    \item \textbf{Edges $\mathcal{E}_{\text{MSG}}$ and Weights:} Directed edges connect adjacent stages only.
        \begin{itemize}
            \item \textit{Computation Edges:} Link $N_{s,k}^{\text{in}}$ to $N_{s,k}^{\text{out}}$ with weight $D^{\text{cp}}_{s,f_k}$, according to Eq. (\ref{eq:comp_delay}).
            \item \textit{Transmission Edges:} Link stage-$k$ out-nodes to stage-$(k+1)$ in-nodes. The weight from satellite $u$ (stage $k$) to satellite $v$ (stage $k+1$) is the pre-computed average delay $\bar{D}^{\text{tx}}(u, v)$. Edges also connect $N_{\text{src}}$ to stage 1, and stage $M$ to $N_{\text{dst}}$ similarly.
        \end{itemize}
\end{itemize}

\subsection{Optimal Path Calculation}
Finding the minimum estimated average delay path in the original problem is equivalent to computing the shortest path from $N_{\text{src}}$ to $N_{\text{dst}}$ in the $\mathcal{G}_{\text{MSG}}$. Since the MSG is a Directed Acyclic Graph (DAG) with non-negative weights, this shortest path can be found efficiently using the topological sorting-based shortest path algorithm for DAGs, with a time complexity linear in the size of the MSG, i.e., $O(|\mathcal{V}_{\text{MSG}}| + |\mathcal{E}_{\text{MSG}}|)$. 
The SA-MSGR algorithm proceeds as follows:
\begin{enumerate}
    \item \textbf{Offline Phase:} Pre-compute and store the average shortest path transmission delays $\bar{D}^{\text{tx}}(u, v)$ for all satellite pairs $(u,v)$. These values are derived from the dynamic network simulation and stability analysis detailed in Section \ref{sec:stability}. This potentially intensive computation is performed once, typically utilizing ground-based resources.
    \item \textbf{Online Phase (executed upon receiving request $r$):}
    \begin{enumerate}
        \item \textit{Construct MSG:} Build the graph $\mathcal{G}_{\text{MSG}}$ tailored to the specific request $r$ (including its VNF sequence $\mathbf{f}$ and source/destination nodes), utilizing the pre-computed $\bar{D}^{\text{tx}}$ for transmission edge weights and the known computation delays $D^{\text{cp}}$ for computation edge weights.
        \item \textit{Find Shortest Path:} Apply dynamic programming to efficiently find the shortest path from the source node $N_{\text{src}}$ to the destination node $N_{\text{dst}}$ within the static DAG structure of $\mathcal{G}_{\text{MSG}}$.
        \item \textit{Output Path:} Extract the sequence of satellites $\pi^*(r)$ corresponding to the computed shortest path, representing the final SFC route.
    \end{enumerate}
\end{enumerate}
The online complexity per SFC request for SA-MSGR is primarily determined by the shortest path calculation on the MSG, roughly $O(M \cdot S_{\text{max}}^2)$ where $S_{\text{max}}$ is the maximum number of satellite choices per VNF stage. It avoids the complexity multiplication associated with the time window dimension in TEG methods, making it significantly more scalable. Compared to the snapshot-based methods, it shifts the expensive shortest path computations on the full LEO network graph to the offline phase, and also avoids the frequent route recalculation at different time slots.

\section{Performance Evaluation}
\label{sec:sim}
In this section, we evaluate the performance of our proposed SA-MSGR algorithm through simulations, focusing on its effectiveness in minimizing the end-to-end SFC delay. We compare SA-MSGR against several baseline routing strategies.

\subsection{Simulation Setup}
\label{subsec:sim_setup}
We simulate a Walker Delta LEO constellation, with $P=12$ planes, $N=30$ satellites per plane, $550$ km altitude, $53^{\circ}$ inclination. Leveraging orbital predictability, the dynamic topology $\mathcal{G}(t)$ and instantaneous ISL delays $d_e(t)$ are pre-computed for one orbital period ($\approx 90$ min, $\Delta t = 1$s), allowing for offline calculation of the average shortest path delays $\bar{D}^{\text{tx}}(u, v)$ used by SA-MSGR. 
Random SFC requests $r=(s_{\text{src}}, s_{\text{dst}}, \mathbf{f})$ are generated with random source/destination nodes. The SFC length $M$ is varied, e.g., $M \in \{5, 10, 15, 20, 25\}$, with the VNF sequence $\mathbf{f}$ comprising $M$ distinct types randomly chosen from $|\mathcal{V}| = 60$ types. VNF complexity $C_{f_k}$ is uniform in $[1, 5]$ GFLOPs, and on-board processor frequency $F_s$ is uniform in $[1, 2]$ GFLOPs per second, yielding computation delays $D^{\text{cp}}_{s,f_k}$.
VNF deployment is randomized per satellite: the number of hosted VNFs $N_s$ follows $\mathcal{N}(3, 0.75^2)$, rounded to the nearest non-negative integer $N_s'$. Then, $N_s'$ distinct VNFs are randomly selected from $\mathcal{V}$ to form the available set $\mathcal{V}(s)$. The primary performance metric is the average joint computation and communication end-to-end SFC delay calculated according to Eq. (\ref{eq:total_inst_delay}). Results are averaged over 300 different SFC requests for each parameter setting. We compare our proposed SA-MSGR algorithm against the following baseline strategies. 
\begin{itemize}
    \item \textbf{Greedy-Transmission (Greedy-Tx)} selects the next hop satellite $s_{k+1}$ hosting $f_{k+1}$ that minimizes the instantaneous shortest path transmission delay $D^{\text{tx}}(s_k, s_{k+1}, t_0)$ from the current node $s_k$.
    \item \textbf{Greedy-Computation (Greedy-Cp)} instead chooses the satellite $s_{k+1}$ hosting $f_{k+1}$ with the lowest computation delay $D^{\text{cp}}_{s_{k+1}, f_{k+1}}$, ignoring transmission costs.
    \item \textbf{Random (Rand)} serves as a basic reference, selecting $s_{k+1}$ uniformly at random from all feasible satellites hosting $f_{k+1}$.
    \item \textbf{Snapshot-based} method optimizes the path using the MSG framework with instantaneous transmission delays $D^{\text{tx}}(u, v, t_0)$.
    \item \textbf{TEG-based} method theoretically optimizes routing over a time window (10 seconds) using a time-expanded graph, which is considered computationally infeasible.
\end{itemize}

\subsection{Results and Analysis}
\label{subsec:results}
\begin{figure}[tbp]
    \centering
    \includegraphics[width=1.0\linewidth]{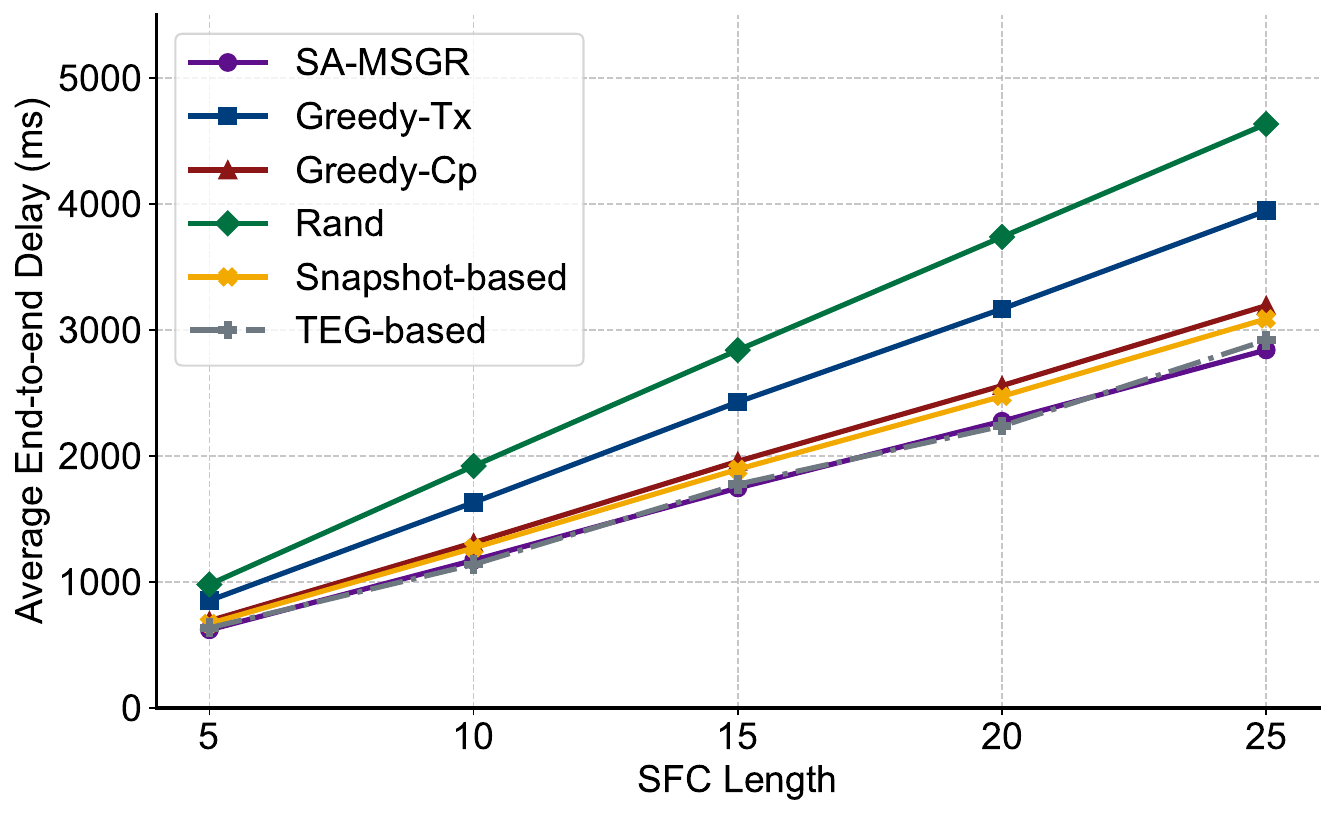}
    \vspace{-7mm}
    \caption{Average end-to-end SFC delay comparison}
    \vspace{-3mm}
    \label{fig:delay_overall}
\end{figure}

As illustrated in Fig. \ref{fig:delay_overall}, SA-MSGR consistently achieves the lowest average end-to-end delay across all tested SFC lengths ($M$). Notably, its performance is remarkably close to the theoretical TEG-based approach, which optimizes over a time window but is computationally infeasible for practical deployment. This suggests that SA-MSGR, by leveraging the statistical stability of path delays, effectively captures the essential long-term characteristics needed for near-optimal average performance without incurring TEG's complexity.
Furthermore, SA-MSGR consistently outperforms the practical snapshot-based method. While the Snapshot-based method optimizes based on the instantaneous network state, relying on potentially unrepresentative transient conditions leads to slightly higher average delays compared to SA-MSGR's stable average-based optimization. Both the SA-MSGR and snapshot-based methods, utilizing the global MSG framework, significantly outperform the simpler heuristics. Consistent with previous observations, Greedy-Cp surpasses Greedy-Tx in average delay in our setup (likely due to computation delays dominating), while Random performs worst. The potential load-balancing issue of Greedy-Cp, however, remains a practical concern not captured by average delay alone.

\begin{figure}[tbp]
    \centering
    \includegraphics[width=1.0\linewidth]{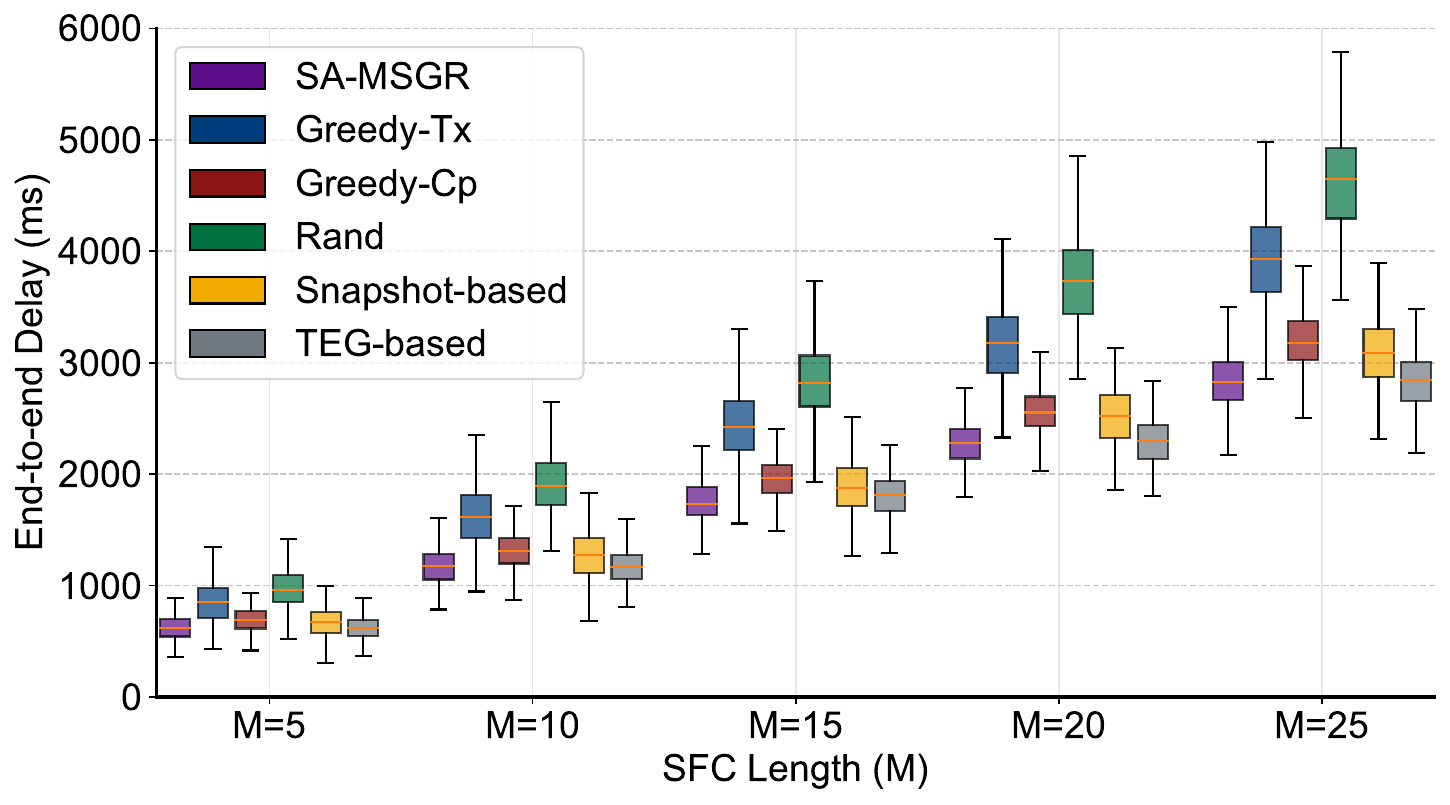} 
    \vspace{-7mm}
    \caption{Distribution of end-to-end SFC delays for different algorithms. The box represents the interquartile range (IQR, 25th to 75th percentile). The line inside the box is the median. Whiskers extend to 1.5 × IQR from the box edges.}
    \vspace{-3mm}
    \label{fig:delay_distribution}
\end{figure}

Beyond averages, Fig. \ref{fig:delay_distribution} reveals crucial differences in delay variability. SA-MSGR shows a highly compact distribution with a narrow interquartile range and short whiskers, indicating consistent and predictable performance, which is desirable for service guarantees.
In stark contrast, the Snapshot-based method exhibits considerably larger delay variance compared to both SA-MSGR and TEG. This highlights a key limitation of snapshot approaches: optimizing based on fluctuating instantaneous conditions leads to greater path diversity and less predictable end-to-end delays. SA-MSGR's use of stable averages mitigates this variability. Among the heuristics, Greedy-Cp displays relatively low variance (possibly due to consistently selecting specific high-power satellites with stable links), whereas Greedy-Tx and Random show the largest variability, reflecting their fluctuating and less predictable routing decision.

\section{Conclusion}
\label{sec:conclusion}
This paper demonstrated statistical stability in LEO network shortest-path delay and proposed the SA-MSGR algorithm, which leverages pre-computed average delays within a MSG framework. Simulations validate that SA-MSGR achieves substantially lower and more consistent delays compared to baselines. Future work could extend this approach to incorporate congestion control and dynamic VNF placement strategies.

\bibliographystyle{IEEEtran}
\bibliography{ref}
\end{document}